\documentclass[aps,showpacs,superscriptaddress,amsfonts,twocolumn,reprint,prl]{revtex4-1} 

\usepackage{graphicx}
\usepackage{array}
\usepackage{hyperref}
\usepackage{amsmath,amssymb}
\usepackage{bm}
\usepackage{float}
\usepackage{amsfonts}
\usepackage{xcolor}

\newcommand{\xfb}{x_{\text{fb}}}
\newcommand{\zfb}{z_{\text{fb}}}
\newcommand{\zf}{z_{\text{f}}}
\newcommand{\vc}{v_{\text{c}}}
\newcommand{\vs}{v_{\text{s}}}
\newcommand{\tc}{t_{\text{c}}}
\newcommand{\rhoF}{\rho_{\text{F}}}

\begin{document}
\title{Light-Control of Localised Photo-Bio-Convection}
\date{\today} 

\author{Jorge Arrieta}
\affiliation{Instituto Mediterr\'aneo de Estudios Avanzados, IMEDEA, UIB-CSIC, Esporles, 07190, Spain}
\author{Marco Polin}
\affiliation{Physics Department and Centre for Mechanochemical Cell Biology, University of Warwick, Gibbet Hill Road, Coventry, CV4 7AL, United Kingdom} 
\author{Ram\'on Saleta-Piersanti}
\affiliation{Instituto Mediterr\'aneo de Estudios Avanzados, IMEDEA, UIB-CSIC, Esporles, 07190, Spain} 
\author{Idan Tuval}
\affiliation{Instituto Mediterr\'aneo de Estudios Avanzados, IMEDEA, UIB-CSIC, Esporles, 07190, Spain}

\begin{abstract}

Microorganismal motility is often characterised by complex responses to environmental physico-chemical stimuli. Although the biological basis of these responses is often not well understood, their exploitation already promises novel avenues to directly control the motion of living active matter at both the individual and collective level.
Here we leverage the phototactic ability of the model microalga {\it Chlamydomonas reinhardtii} to precisely control the timing and position of localised cell photo-accumulation, leading to the controlled development of isolated bioconvective plumes. 
This novel form of photo-bio-convection allows a precise, fast and reconfigurable control of the spatio-temporal dynamics of the instability and the ensuing global recirculation, which can be activated and stopped in real time. 
A simple continuum model accounts for the phototactic response of the suspension and demonstrates how the spatio-temporal dynamics of the illumination field can be used as a simple external switch to produce efficient bio-mixing.
\end{abstract}

\maketitle

The autonomous movement of microorganisms has fascinated scientists since the discovery of the microbial world. Particularly striking is the variety of coordinated collective dynamics that emerges with startling reliability in groups of motile microorganisms, from traffic lanes and oscillations in bacterial swarms \cite{Chen2017,Ariel2013}, to wolf-pack hunting \cite{Berleman2009} and microbial morphogenesis \cite{Chisholm2004,Deng2014}. Nowadays, microbial motility is an important part of a growing interdisciplinary field aiming to uncover the fundamental laws governing the dynamics of so-called active matter \cite{Marchetti2013},  eventually allowing us to harness micron-scale motility for applications ranging from targeted payload delivery \cite{Koumakis2013} to direct assembly of materials \cite{Angelani2011,Ma2017},
 with either living organisms or synthetic microswimmers \cite{Kummel2015,Bechinger2016,Aubret2018}.
Realising this potential will hinge on our ability to alter and ultimately control the motion of both individual cells and microbial collectives. 

Microbial motility can be controlled through clever engineering of  boundaries 
\cite{Denissenko2012,Wioland2016,Thutupalli2017} 
or topology \cite{Giomi2016,Morin2018} of the vessels holding the microbial suspension, leading for example to  predictable  accumulation \cite{Galajda2007,Kantsler2013,Ostapenko2016} or circulation of cells \cite{Wioland2016,Lushi2014,Bricard2013,Vizsnyiczai2017}. 
However, together with strategies squarely rooted in physics, control of living microswimmers can rely also on approaches bridging between physics and biology, by taking advantage of pathways linking motility with the perception of physico-chemical stimuli by cells. 
Light is particularly well suited to this end: its manipulation is readily achievable at both the macroscopic and microscopic  \cite{Lee2010,Stellinga2018} scales; and it is an important stimulus for a wide variety of microorganisms, providing both energy  \cite{Dodd2013} and information often used to prevent potentially lethal light-induced stress \cite{Li2009b}.
Most microorganisms respond to light by linking swimming speed to light intensity (photokinesis, \cite{Hader1987}) and/or re-directing their motion towards or away from the light source (phototaxis, \cite{Hader2009,Drescher2010,Arrietaetal2017}).  
Although the physiological details underpinning these active responses are often not completely understood \cite{Arrietaetal2017,Giometto2015,Leptos2018,Drescher2010}, techniques employing light to precisely control the dynamics of swimming microorganisms are already emerging. 
Biological responses to light led to the development of genetically-engineered light-sensitive bacteria \cite{Arlt2018,Jin2018,Huang2018} used, for example, to power micron-size motors \cite{Vizsnyiczai2017}, and have even inspired the fabrication of light-reactive artificial swimmers \cite{Maggi2015,Bechinger2016,Lozano2016,Dai2016}.

\begin{figure}[b!]
\centering
\includegraphics[width=\columnwidth]{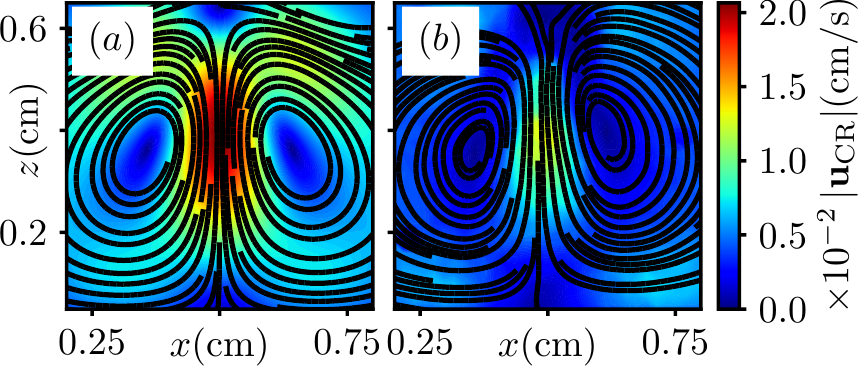}
\caption{Contours of the cells averaged velocity field obtained from $(a)$ the numerical integration of~\eqref{eq1}--\eqref{eq3} for $n_0=1.5\times10^7\mathrm{cells}/ml$ and $\beta=0.14$ and $(b)$ from the PIV analysis of experimental data. Solid black lines represent in both panels the corresponding streamlines.}
\label{fig:plume_streamlines}
\end{figure}

\begin{figure*}[t!]
\begin{center}
\includegraphics*[width=1.8\columnwidth]{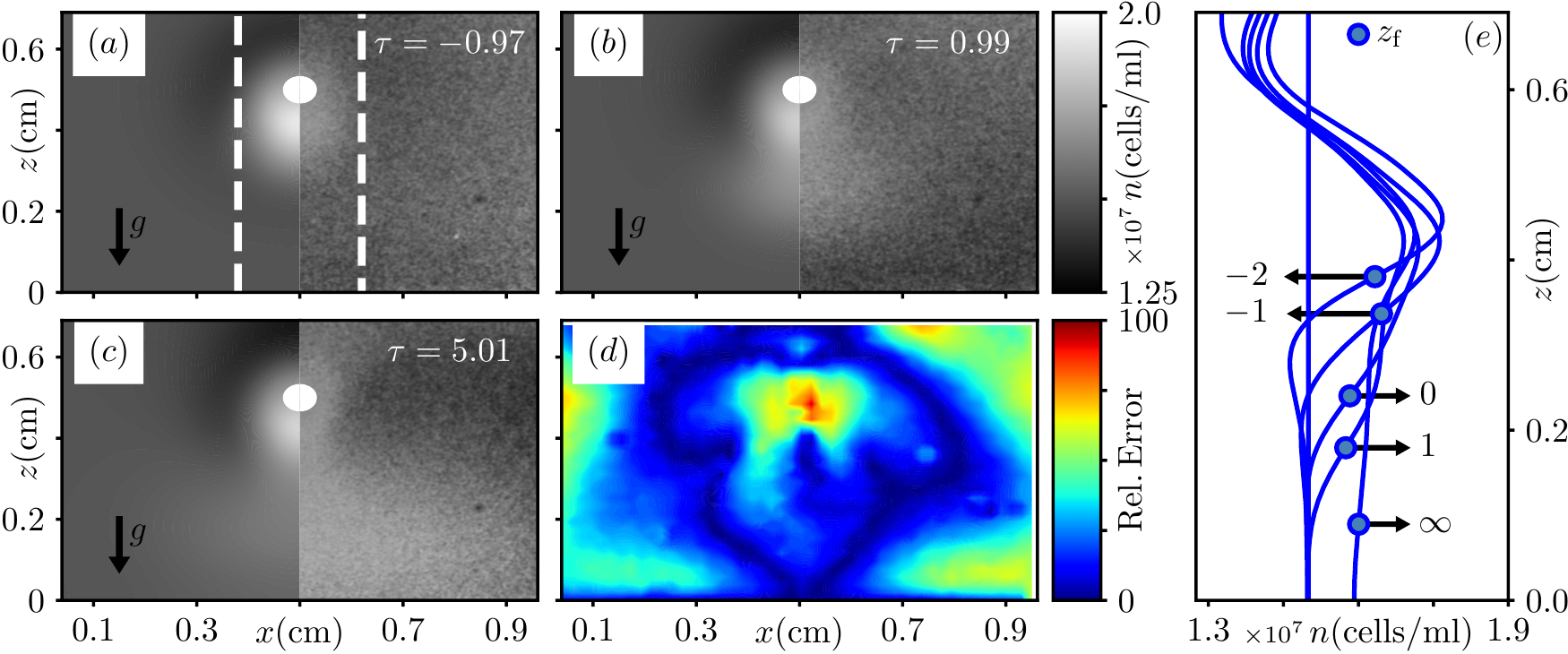}
\end{center}
\caption{Dynamics of plume formation. (a-c) Evolution of photo-accumulation and plume formation at different values of the normalized time $\tau$. Panels show the cell density from the continuum model, $n$ (left-side), and the dark-field image from the corresponding experiment at the same value of $\tau$. White dots in (a-c) correspond with the position of the light fibre. (d) Relative error between experimental and numerical flows of cells  for the steady state flow (each flow field has been first rescaled by its maximum value).(e) Average vertical cell concentration profiles for the specified values of $\tau$, and corresponding values of $\zf$ (circles).}
\label{fig:photoaccumulation}
\end{figure*}

Within eukaryotes together with recent studies on  {\it Euglena gracilis} (Protozoa) \cite{Ozasa2011,Shoji2014,Lam2017,Tsang2018},  current applications focus in particular on the model unicellular green alga {\it Chlamydomonas reinhardtii} (CR), and range from micro-cargo delivery by individual cells  \cite{Whitesides2005} to trapping of passive colloids by light-induced hydrodynamic tweezers \cite{Dervaux2016}, and photofocussing of algal suspensions through an interplay of photo- and gyro-taxis~\cite{Garcia2013}. 
Unlocking the full potential of light-based control, however, will require the development of techniques based on a collective response that is both quick and localised. Despite considerable progress, this is not currently available.

Here we exploit the phototactic response of CR to demonstrate a novel form of dynamic control of a cell suspension, based on a fast ($\sim10\,$s) accumulation that can be localised anywhere within the suspension. %
Cells photo-accumulate around the light from a horizontal optical fibre (Fig.~\ref{fig:plume_streamlines}b) and act as a miniaturised pump driving a global recirculation of the suspension with a fast response time, quantitatively captured by a simple model (Fig.~\ref{fig:plume_streamlines}a).
The fast response of the suspension can be exploited for efficient bio-mixing, an attractive solution to improve current photo-bio-reactor technology for biofuel production where mixing is essential to distribute nutrients, and transfer gases across gas-liquid interfaces \cite{Borowitzka1999,Greenwell2010,Scott2010}. Our results serve as a proof-of-principle for more complex instances of light-controlled fluid flows in biological suspensions.

Unicellular biflagellate green algae  \textit{Chlamydomonas reinhardtii} wild type strain CC125  were grown axenically at $20^{\circ}\,$C in Tris-Acetate-Phosphate medium (TAP) \cite{Harris2009} under fluorescent light illumination (OSRAM Fluora, $100\,\mu$mol/m$^2$s PAR) following a $14\,$h/$10\,$h light/dark diurnal cycle. Exponentially growing cells were harvested, photo-accumulated, diluted to the target concentration with fresh TAP, and  loaded in a vertical observation chamber formed by a square shaped Agar-TAP gasket of $L=1\,$cm side and $h=1\,$mm thickness,  sandwiched between two coverslips. The main experiments were performed at two average concentrations: $n_0^{\text{h}}=1.5\times10^7\,$cells/ml ($8$ repeats); and $n_0^{\text{l}}=7\times10^6\,$cells/ml ($6$ repeats). Tests for plume formation were also conducted at $n_0 = 6.1,9.0\times10^6; 1.24,1.40\times10^7\,$cells/ml.
The suspension's dynamics was visualised through darkfield illumination at $635\,$nm (FLDR-i70A-R24, Falcon Lighting Germany) and recorded by a CCD camera (Pike, AVT USA) hosted on a continuously focusable objective (InfiniVar CFM-2S, Infinity USA). 
Localised actinic illumination was provided by a $200\,\mu$m-diameter horizontal multimode optical fibre (FT200EMT, Thorlabs USA) coupled to a $470\,$nm high-power LED (M470L2, Thorlabs USA). The fibre's output intensity $I(\mathbf{x})$, centred at $\mathbf{x}_{\text{fb}}=(\xfb,\zfb)$ (Fig.~\ref{fig:photoaccumulation}a), is well approximated by the Gaussian used in numerical simulations throughout the manuscript (width $\sigma_I = 667 \mu$m; peak intensity $I_0=260\,\mu$mol/m$^2$s; Fig.~S1\cite{supplementary}).

Figure~\ref{fig:photoaccumulation} shows the evolution of the photo-accumulation dynamics for $n_0^{\text{h}}$. Without light stimuli, individual cells swim in a characteristic run-and-tumble-like behaviour \cite{Polin2009a} leading  to a uniform spatial distribution at the population level (Fig.~\ref{fig:photoaccumulation}a). 
As the actinic light is switched on, phototactic cells start accumulating around the fibre, through a characteristic phototactic steering mechanism \cite{Leptos2018} based on an interplay between time-dependent stimulation of a light-sensitive organelle \cite{Foster1980,Kateriya2004a} and the ensuing flagellar response \cite{Josef2006}. Phototaxis leads, within $\sim 10\,$s, to a $\sim2\,$mm-wide region of high cell concentration \cite{Arrietaetal2017}. This is gravitationally unstable, and eventually falls forming a single, localised sinking plume of effectively denser fluid (Fig.~\ref{fig:photoaccumulation}b-d and Supplementary Movie S1). The system converges to its steady state as the plume reaches the bottom of the container ($\sim 30\,$s), two orders of magnitude faster than reported for alternative configurations~\cite{Dervaux2016} with the cells advected along the strong global recirculation seen in Fig.~\ref{fig:plume_streamlines}a (experimental flow of cells obtained with Open-PIV using cells as tracers~\cite{Tayloretal2010} ). This buoyancy-driven instability is reminiscent of bioconvection, one of the best known collective phenomena in suspensions of microswimmers \cite{Childress1975,PedleyKessler1992,Bees1998}. 
Here it can be understood as a light-induced instance of a single bioconvective plume, which can be actively modulated by light and localised anywhere within the sample. 
Cell accumulation, however, does not always lead to plumes. In samples with average concentration $n_0^{\text{l}}$, the photo-accumulated high-concentration region does not sink to the bottom but reaches instead a stable height just below the fibre's centre. Despite the absence of a proper plume, however, the background fluid is still globally stirred (Supplementary Movie S2).
\begin{figure}[!t]
	\includegraphics[width=0.9\columnwidth]{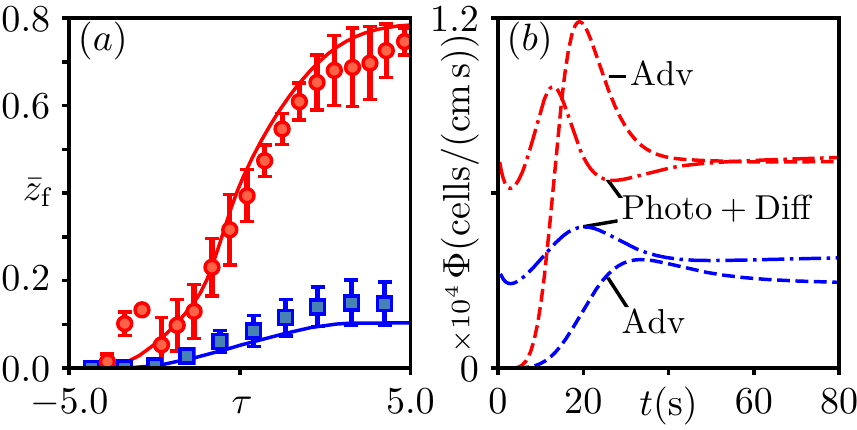}
	\caption{Plume falling and flux balance. $(a)$ Time evolution of the experimental (circles/squares) and numerical (solid lines) fronts vs. $\tau$, for $\zfb\sim0.5\,\mathrm{cm}$. Simulations were run with $n_0=1.5\times10^{7}\,\mathrm{cells/ml}$ (red line) and $n_0=7.0\times10^{6}\,\mathrm{cells/ml}$ (blue line); other parameters as described in the text. $(b)$ Evolution of the advective (dashed lines) and phototactic plus diffusive (dot-dashed lines) fluxes across a circular control surface of radius $0.18\,$cm for the two cases described in panel $(a)$. Same colour code.}
	\label{fig:front_dynamics}
\end{figure} 

Sinking of the initial region of photo-accumulated cells can be quantified from the average vertical profile of the recorded images within a $2\,$mm-wide strip around $\xfb$ (Fig.~\ref{fig:photoaccumulation}a).  The position of the profile's maximal vertical derivative, $\zf$, provides a faithful measure of the height of the photo-accumulated front, which is easy to follow in time (Fig.~\ref{fig:photoaccumulation}e). A heuristic description of its dynamics through the sigmoid function $\zf(t)-\zf(0)= \zf^{\infty}\exp[(t-t_0)/\delta]/\left(\exp[(t-t_0)/\delta]+1\right))$ can be used for both temporal registration, through a parameter to set a common origin of time $t_0$ ; and rescaling, by the characteristic falling time $\delta$. Typically, $\delta=15.5\pm6.6\,$s for $n_0^{\text{h}}$, and $2.4\pm 1.1\,$s for $n_0^{\text{l}}$ (errors are standard deviations of measurement sets). Figure~\ref{fig:front_dynamics}a shows the average rescaled front dynamics $\bar{z}_{\text{f}}(\tau)=(\zf(\tau)-\zf(0))/\zf(0)$ in terms of the intrinsic time $\tau = (t-t_0)/\delta$. The front falls almost to the bottom of the sample ($\bar{z}_{\text{f}}=1$) in the high concentration case ($n_0^{\text{h}}$ blue circles) while  in the low concentration case ($n_0^{\text{l}}$ red circles) the steady-state position is just $\sim1\,$mm below the fibre ($\bar{z}_{\text{f}}\simeq0.1$). This hints at the existence of a bifurcation between $n_0^{\text{l}}$ to $n_0^{\text{h}}$.

The system's behaviour, and  the bifurcation, can be rationalised through a simple continuum model of 2D photo-bioconvection. 
The model describes  the coupling between the local cell density, $n(\mathbf{x},t)$, and the fluid flow, $\mathbf{u}(\mathbf{x},t)$ $\left(\mathbf{x} = (x,z)\right)$. The former obeys a continuity equation which includes contributions from the cells' active diffusion, phototaxis, and advection by the local background flow.  The latter follows the Navier-Stokes equations, coupled to $n(\mathbf{x},t)$ through the cells' excess density ($\Delta\rho$) over the surrounding fluid (density $\rhoF$). Following previous work \cite{Childress1975,PedleyKessler1992} this  is captured in the Bousinnesq approximation.
This minimal model recapitulates well the emergence and falling dynamics of a plume, and the geometric structure of the ensuing recirculation. Therefore, in keeping with a minimal-model approach, we will not consider gravitaxis \cite{Childress1975}, gyrotaxis \cite{PedleyKessler1990}, and the effect of cells' activity in both the bulk stress and the cell diffusivity tensors \cite{PedleyKessler1990, PedleyKessler1992}, despite their role in phenomena like spontaneous bioconvection \cite{PedleyKessler1990,PedleyKessler1992,Bees1998} and cells' focussing \cite{Garcia2013}.  We note, however, that they could still contribute to a global rescaling of the dynamics. 
The system, contained within a square cavity of side $L$, is described by the following set of equations:
\begin{align}
\nabla\cdot \mathbf{u}&=0,\label{eq1}\\
\frac{\partial\mathbf{u}}{\partial t}+\mathbf{u}\cdot\nabla\mathbf{u}&=-\nabla p-n\mathbf{\hat{z}}+\sqrt{\frac{Sc}{Ra}}\nabla^2\mathbf{u},\label{eq2}\\
\frac{\partial n}{\partial t}+\nabla\cdot\left[\left(\mathbf{u}+\mathbf{u}_{\text{ph}}\right)n\right]&=\frac{1}{\sqrt{Ra\,Sc}}\nabla^2 n,
\label{eq3}
\end{align}  
with no-slip at the boundary, and no cell flux through the boundary. 
Despite using the same symbols for convenience, Eqs.~(\ref{eq1}-\ref{eq3}) have been non-dimensionalised by using $L$ as the characteristic length, and introducing the characteristic velocity for buoyancy-driven flow, $\vc=(n_0g\Delta\rho V_{\text{CR}}L/\rhoF)^{1/2}$, where $V_{\text{CR}}=4\pi r_{\text{CR}}^3/3$ is the estimated volume of an individual cell assuming a sphere of radius $r_{\text{CR}}$. The characteristic time is then $\tc=L/\vc$; the scale for the (2D) pressure $p$, is given by $h\rhoF\vc^2$; and $n_0$ rescales the cell density.  The behaviour of the system is dictated by three non-dimensional numbers: a Rayleigh number, $Ra=(\vc L)^2/\nu D$, and a Schmidt number, $Sc=\nu/D$, based on the kinematic viscosity of the fluid ($\nu$) and the cells' effective diffusivity ($D$); and the phototactic sensitivity $\beta$, which governs the non-dimensional phototactic term, $\mathbf{u}_{\text{ph}}=\beta (\vs/\vc) (|\mathbf{x}-\mathbf{x}_{\text{fb}}|/h^*)\nabla I/|\nabla I|_{\text{max}}$. 
The phototactic drift, derived and tested in \cite{Arrietaetal2017}, includes the cells' swimming speed, $\vs = 7.8\times10^{-3}\,$cm/s, and the effective  thickness of the illuminated chamber, $h^*=5.19\times10^{-2}\,$cm. The experimental system is expected to correspond to $\beta=0.14$ \cite{Arrietaetal2017}. Other parameters are fixed to:  $\Delta\rho=0.05\,$g/cm$^3$; $\rhoF=1\,$g/cm$^3$; $\nu=10^{-2}\,$cm$^2$/s; $r_{\text{CR}}=5\times10^{-4}\,$cm; $D=3.9\times10^{-4}\,$cm$^2$/s \cite{Polin2009a,Arrietaetal2017}.
\begin{figure}[!t]
	\includegraphics[width=0.9\columnwidth]{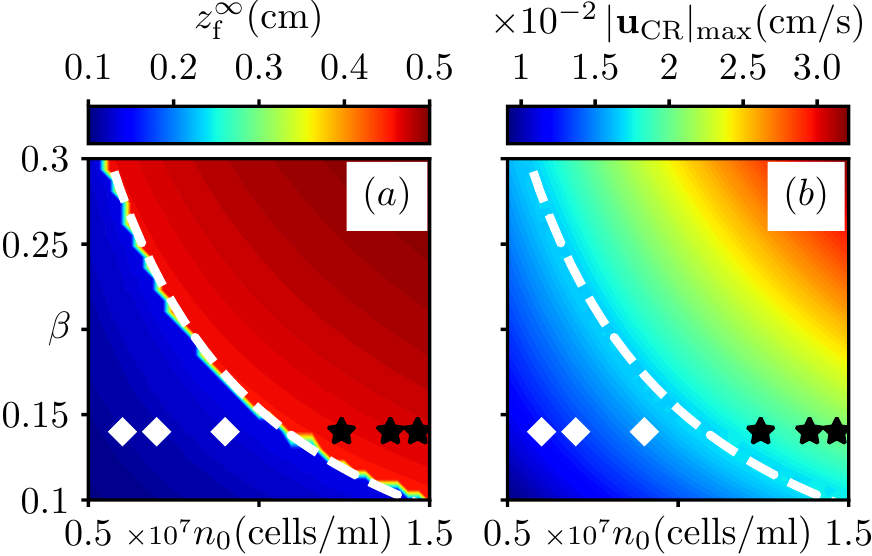}
	\caption{Photo-bio-convective phase behaviour.  $(a)$ Final position of the plume front, $\zf^\infty$ and, $(b)$ maximum velocity of the cells, $|\mathbf{u}_{\text{CR}}|_{\text{max}}$, vs. initial cell density $n_0$ and photo-adaptation parameter $\beta$. Plume forming/non-forming experiments are indicated by black squares/white diamonds respectively.  The isoline $|\mathbf{u}_{\text{CR}}|_{\text{max}}=1.66\time 10^{-2}$cm/s separates not-falling/falling regimes (white dashed line).}
	\label{fig:phase_space}
\end{figure}
Cells are initially uniformly distributed within the quiescent fluid, and the vorticity-stream function formulations of Eqs.~(\ref{eq1}-\ref{eq3}) are integrated with a spatially centred, second-order accurate, finite-difference scheme~\cite{Anderson1984}; at each intermediate stage of the third-order Runge-Kutta method used to advance time, the Laplace equation for the stream function is solved with the conjugate gradient method~\cite{Ferziger2001}. The integration scheme was validated with benchmark solutions~\cite{Davis1983}.

Figure~\ref{fig:photoaccumulation} compares experimental and numerical dynamics of plume formation and sinking, as a function of the reduced time $\tau$ ($\zfb=0.5\,$cm; $n_0=n_0^{\text{h}}$). The agreement is excellent with no fitting parameters, and it is maintained also at longer times. Figure~\ref{fig:photoaccumulation}d shows the relative error between cells' stationary velocity field $\left(\mathbf{u}_{\text{CR}}=(\mathbf{u}+\mathbf{u}_{\text{ph}})\vc\right)$ from experiments and model, rescaled by their peak velocity. The small discrepancy ($<25\%$ on average) shows that the model captured well the structure of the photo-bioconvective flow of cells (see also Fig.~\ref{fig:plume_streamlines}).
A closeup, in Fig.~\ref{fig:front_dynamics}, on the experimental (circles) and numerical (solid lines) front dynamics, proves that the model captures both sinking and non-sinking regimes (respectively $n_0^{\text{h}}$ and $n_0^{\text{l}}$). We therefore decided to explore systematically the system's behaviour {\it in silico} through a parametric sweep in the range $n_0\in[0.5,1.5]\times10^7\,$cells/ml and $\beta\in[0.1,0.3]$.
Figure~\ref{fig:phase_space}a,b shows that both the characteristic falling time $\left(\delta\right)$ and the steady-state front position $\left(\zf^\infty\right)$ indicate the presence of two distinct regimes separated by a sharp transition, in line with experiments (Fig.~\ref{fig:phase_space}, star marks). There is a low-$n_0$/low-$\beta$ regime, where cells accumulate but do not fall; and a high-$n_0$/high-$\beta$ one, where the light induces a single, isolated bioconvective plume driving a vigorous global recirculation (Fig.~\ref{fig:plume_streamlines}). 
The two regimes are separated by a critical curve $\beta n_0\approx~$constant, corresponding to the isoline of maximum cells' velocity, $|\mathbf{u}_{\text{CR}}|_{\text{max}}\simeq 1.66\times10^{-2}\,$cm/s. This is compatible with the full set of $n_0$ values explored experimentally (Fig.~\ref{fig:phase_space}b).
The process driving the bifurcation can be understood intuitively by examining the balance between phototactic, diffusive and advective fluxes of cells. Figure~\ref{fig:front_dynamics}b shows the evolution of these fluxes across a circle of radius $18\,$mm centred on the optical fibre. Before the bifurcation (blue) advection (dot-dashed line) is always lower than the net flux due to cell motion (phototaxis and diffusion, dashed line): the front remains close to its initial position. Beyond the bifurcation (red) the downward advective flux dominates shortly after the initial accumulation, transporting cells downwards with a critical velocity arising from flux balance.

 \begin{figure}[!t]
 	\begin{center}
 		\includegraphics*[width=0.9\columnwidth]{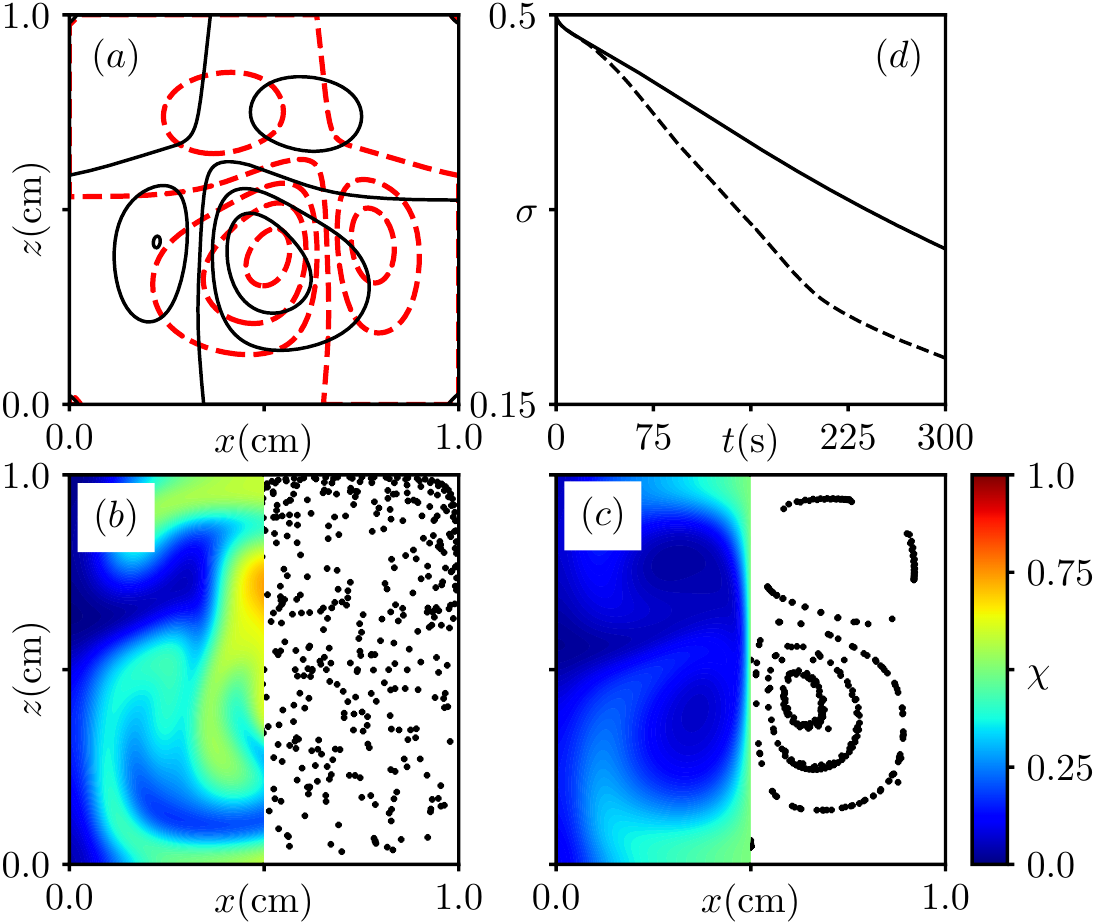}
 		\caption{Mixing by blinking plumes. $(a)$ Streamlines of the induced flow at half a period (black line) and a full period (red line) of the time-dependent light protocol. (b,c) Spatial distribution of $\chi$ (left-hand side) at $t\simeq200\,$s and Poincar\'e map (right-hand side) after $i=75$ periods for the ``blinking plumes" and the centred stationary case, respectively. $(d)$ Time evolution of the spatial standard deviation of $\chi$ computed in $(b)$ (dashed line) and $(c)$ (solid line).}
 		\label{fig:mixing}
 	\end{center}
 \end{figure}
The ability to determine location and timing of the plume formation can be harnessed to govern the global transport properties of the suspension, e.g. to accelerate the active biomixing of nutrients.
A simple procedure takes advantage of the left-right asymmetric flows generated when the light source is shifted from the mid-point of the chamber, as shown in the streamlines of Fig.~\ref{fig:mixing}a for a $\pm1.5\,$mm shift (dashed red and solid black lines respectively), and experimentally in Supplementary Movie S3. Alternating evenly between the two plumes in a cycle of period $T$, generates flow fields that display the characteristic crossing of streamlines required for efficient mixing in 2D, and realise within a photo-bioconvective context the blinking vortex, a paradigmatic example of mixing by chaotic advection \cite{Aref2017}.
Figure~\ref{fig:mixing}b,c and Supplementary Movie S4 show how the concentration $\chi$ of an advected nutrient of diffusivity $D_\chi=3\times10^{-3}\,$mm$^2$/s, mimicking photosynthetically-important gases like CO$_2$ \cite{Mazarei1980}, evolves from an initial distribution localised in the right half of the container, for $t\sim 200\,$s.
Crossing of streamlines leads to the stretching and folding of thin filaments characteristic of chaotic advection. These in turn cause a significantly faster mixing than for a single steady plume (Fig.~\ref{fig:mixing}d), as seen in the decay of the standard deviation of the spatial concentration profile,  $\sigma^2 = \langle(\chi-\langle \chi\rangle)^2\rangle$ (Fig.~\ref{fig:mixing}e) \cite{Stroock2002}. The origin of the enhanced mixing is evident in the Poincar\'e maps obtained from the trajectories of ten tracer particles  initially distributed uniformly along $z=5\,$mm and followed over $75\,T$ (Fig.~\ref{fig:mixing}c,d right panels).
Closed quasi-periodic orbits are readily visible for a stationary centred single plume while the ``blinking plumes'' lead to particles exploring most of the spatial domain. Active light patterning can therefore induce mixing advective maps, leading to strongly enhanced nutrient transport throughout the cell culture.

 We have presented a novel mechanism that harnesses phototaxis to actively control a suspensions of swimming microorganisms through their accumulation around a localised light source. The ensuing global instability, characterised by steady vortical flows for all parameter values, can easily lead to the emergence of isolated bioconvective plumes whose spatio-temporal localisation is simply tuned by the external illumination. These properties contrast with the limited control afforded by standard bioconvection 
\cite{Vincent1996, Williams2011, Williams2011a,Shoji2014, Panda2016}, 
and enable rapid light-mediated control of the flow which can be used to regulate the transport properties of the suspension. 
The simple minimal model we use provides a surprisingly accurate quantitative description of the experimental system with no fitting parameters, but only when viewed in terms of the reduced time $\tau$. 
In terms of real time, plumes fall slower in experiments than in simulations: $\delta=10.5\pm 4.7\,$s across all experiments beyond the bifurcation, compared to a range $4.5\,\text{s}-4.8\,\text{s}$ expected from the model. Interestingly, the quantitative agreement is much improved below the bifurcation (exp: $\delta=4.8\pm 2.5\,$s; num: $3.1\,\text{s}-3.7\,\text{s}$).
The discrepancy beyond the bifurcation is possibly coming from a combination of disregarded swimming features (gravitaxis and gyrotaxis \cite{Childress1975,PedleyKessler1990}) and confinement \cite{Pushkin2016},  to be disentangled in a future, dedicated study.
Overall, together with recent work pioneering the use of radial stresses~\cite{Dervaux2016}, our results set the stage to use light for fast and complex spatio-temporal control of the macroscopic dynamics of phototactic suspensions.

\begin{acknowledgments}
We acknowledge the support of the Spanish Ministry of Economy and Competitiveness Grants No. FIS2016-77692-C2-1-P (IT) and CTM-2017-83774-D (JA), and the subprogram Juan de la Cierva No. IJCI-2015-26955 (JA). JA is extremely grateful to Sara Guerrero for her enormous encouragement and support in the development of this work. MP and IT would like to thank Raymond Goldstein for support in the initial stages of the project. JA is grateful to Rapha\"{e}l Jeanneret for thoughtful discussions.
\end{acknowledgments}

\bibliographystyle{aipnum4-1}
%

\end{document}